    \documentclass[11pt,letterpaper]{article}
    \usepackage{mathlet}
    \renewcommand{\cal}{\mathcal}
\font\titlfnt = cmss17
\def\gtrsim {\mkern 5mu\raise .425ex\hbox{$>$}\mkern -14mu%
\lower .7ex\hbox{$\sim$}\mkern 5mu}
\def\lesssim {\mkern 5mu\raise .425ex\hbox{$<$}\mkern -13.3mu%
\lower .7ex\hbox{$\sim$}\mkern 5mu}
\def\slsh#1{#1\!\!\! /}

\newcounter{insertionEq}
\newcounter{FlatTmunuEq}
\newcounter{hatEq}
\newcounter{insertionFinalEq}
\newcounter{selfcorrectedEq}
\newcounter{HGandBEq}
\newcounter{hGandBLightConeEq}
\newcounter{vanishingFlatDivergenceEq}
\newcounter{fofZEq}
\newcounter{ZsimEq}
\newcounter{HofxEq}
\newcounter{hsimEq}
\newcounter{HsimEq}
\newcounter{QFTmunuEq}
\newcounter{DEq}
\newcounter{TmunuCurvedEq}

\begin{document}

\bibliographystyle{unsrt}

\renewcommand{\thesection}{\Roman{section}}
\renewcommand{\thefootnote}{\fnsymbol{footnote}}

\begin{titlepage}

\rightline{hep-th/9505191}\par

\vspace{7.5mm}

\begin{center}

{\titlfnt
Self-Gravitational Correction of the
``\thinspace Vacuum Polarization''
Feynman Diagram Using Full Einstein
Equation Propagation of
the Intermediate Virtual Gravitons}
\vspace{10mm}

\noindent
{\large
S. K. Kauffmann}\footnote{
E-Mail address:
74631.3052@compuserve.com\par
\ \ \hskip 0.1em Telephone: 1(415)585-0678}\\
750 Gonzalez Drive, Apt.\ 6D, San Francisco, CA 94132-2208, U.S.A.

\vspace{5mm}

\end{center}

\vspace{10mm}

\begin{abstract}

The second-order ``vacuum polarization'' radiative
correction insertion Feynman
diagram is an ultra\-violet divergent virtual object
whose Fourier transform to configuration
space--time has the properties of a virtual
stress-energy tensor.
Though well-defined, this virtual stress-energy is
non-integrably singular on the
light cone---which produces the aforementioned
ultraviolet divergence of its
four-momentum space Fourier transform.
To properly calculate the self-gravitationally
corrected version of this singular virtual
stress-energy, the usual graviton-exchange
ladder sum approximations are eschewed
in favor of full, self-consistent Einstein
equation propagation of the intermediate
virtual gravitons, which takes into account
their important non-linear interactions with
each other.  (As a by-product, the subsequent
perturbative treatment of these non-linearities
is avoided, which eliminates the source of the
ultraviolet divergences of the second-quantized
gravity theory itself.)  The resulting
corrected virtual stress-energy is non-singular
everywhere and Fourier-transforms convergently
to a finite corrected version of the diagram.
This corrected diagram makes no
contribution to charge renormalization
(as could be expected of a diagram
involving but a single transient virtual pair),
and its dynamical behaviour
accords with the standard
quantum electrodynamics result except at inaccessibly
extreme values of the momentum transfer, $|q^{2}|\gtrsim
(Ge^{2})^{-1}$.  There, the standard
logarithmic rise with momentum
transfer which this diagram
contributes to the effective coupling
strength falls away, as the diagram
proceeds instead to decrease strongly
toward zero.  The same self-gravitational
correction is made to the closely related
quartically divergent second-order
vacuum-to-vacuum amplitude correction
Feynman diagram, and it is found that
the result vanishes identically.

\vspace{5mm}
\end{abstract}

\vspace{37.5mm}

\noindent
May 1995

\end{titlepage}

\renewcommand{\thefootnote}{\arabic{footnote}}
\addtocounter{footnote}{-\value{footnote}}

%%%%%%%%%%%%%%%%%%%%%%%%%%%%%%%%%%%%%%%%%%%%%%%%%%%%%%%%%%%%%%%%%%%

\section{Introduction}
\label{Sect:Introduction}
\addtocounter{equation}{-\value{equation}}

Self-gravitational correction as a possible
physical mechanism for resolving
ultraviolet divergences in quantum field theories
has been discussed over
the years by a number of authors.
Some of the earliest published speculations
along this line were put forth by
Landau
\cite{Landau},
Klein
\cite{Klein1,Klein2,Klein3},
Pauli \cite{Pauli},
and Deser
\cite{Deser}.  A definitive determination
of the finite mass of the classical
point charge was worked out by
Arnowitt, Deser, and Misner
\cite{ADM} in the context
of a non-singular metric (different from the conventional singular
Reissner--Nordstr{\"o}m metric for this situation).
This mass for the classical
point charge has recently been
independently rederived by the author \cite{Kauffmann}.
Ref.~\cite{ADM} emphasizes that, for
this highly singular source of gravitational
field, the countervailing (negative)\
effective energy density due to
the resulting very strong
gravitational field itself,
carefully taken into account in
self-consistent fashion,
is critical to obtaining the correct finite
value for the point charge's mass.
Subsequent work by DeWitt \cite{DeWitt},
Khriplovich \cite{Khriplovich},
Salam and Strathdee \cite{SS},
and Isham, Salam, and Strathdee \cite{ISS}
has approached the ultraviolet
divergences of quantum
electrodynamics through correcting the
lowest-order Feynman diagrams in which they
occur by various types of summed graviton-exchange
ladders.  These simple graviton-exchange
processes do not, however, include the non-linear
feature that the virtual gravitons interact
{\em with each other}, which was
found in Ref.~\cite{ADM}
to be so crucial to the proper
self-gravitational resolution of the infinite
mass of the classical point charge.
Moreover, the graviton-exchange ladder
sums are generally not by themselves
gravitationally gauge-invariant \cite{DeWitt,Khriplovich},
and can in addition introduce difficulties with
electromagnetic gauge invariance \cite{ISS}.
Khriplovich looked into the effects of making
different choices of gravitational gauge in his
approach, and found that some of these seemed
unlikely to result in the suppression of the
ultraviolet divergences \cite{Khriplovich2}.
Furthermore, the quantized electron mass
electromagnetic correction obtained in
Ref.~\cite{Kauffmann} (by straightforward
extension of the non-singular metric
approach that was used for the classical
point charge)\ is found to incorporate a
gravitationally induced effective cutoff
radius which depends in a physically
sensible way on the electron charge---in
contrast with the puzzling charge independence
of that effective cutoff radius in the result
obtained by means of the ladder-sum approaches
of Refs.~\cite{Khriplovich} and \cite{ISS}.

The self-gravitational treatment in this paper
of the ultraviolet-divergent
second-order ``vacuum polarization''
radiative correction insertion Feynman diagram of
Fig.~\ref{Fig:VacPol} is essentially the same as the quantum
field theoretic approach of Ref.~\cite{ISS} in
most respects.  Since the focus here is on
self-gravitational suppression of the quantum
electrodynamically induced ultraviolet divergence,
the Ref.~\cite{ISS} restriction that gravity is
only coupled to the electromagnetic interaction term of
the action integral is observed, as it
is this interaction term which generates the Feynman
diagram electromagnetic vertices, without which
there could be no electromagnetically induced
ultraviolet divergences.  Thus the diagram of
Fig.~\ref{Fig:VacPol} is permitted to couple to
intermediate virtual gravitons which travel between
its two vertices---these produce the part of its
self-gravitational correction which could be capable
of impinging on its ultraviolet divergence.
Ref.~\cite{ISS} uses the ``graviton superpropagator''
approximation framework to describe the propagation
of these intermediate virtual gravitons.  This
``graviton superpropagator'' is, of course, a member
of the species of summed graviton-exchange
ladders---whose shortcomings in describing ultrastrong
gravitational effects engendered by an ultraviolet
divergence were touched upon in the discussion above.
The ``superpropagator'' approximation of Ref.~\cite{ISS}
is therefore replaced by use of the full Einstein field
equation to self-consistently describe the propagation of
the intermediate virtual gravitons.  Fourier transformed
from four-momentum space to configuration space--time,
the insertion diagram of Fig.~\ref{Fig:VacPol} has the proper
dimensions and other attributes of a stress-energy tensor,
although, being complex-valued, it is, of course, a virtual
stress-energy.  It also possesses a non-integrable
singularity on the light cone, which is the feature
that results in its ultraviolet divergence upon Fourier
transformation back to four-momentum space.  This
complex-valued virtual stress-energy gives rise, via
the Einstein equation, to a complex-valued virtual
metric that far more satisfactorily describes the
intermediate virtual graviton propagation in this
ultraviolet-divergent situation than does the
``superpropagator'' approximation of Ref.~\cite{ISS}.
{}From this virtual metric we obtain the
self-gravitationally corrected virtual stress-energy
by the method of Ref.~\cite{Weinberg},
namely we calculate $-(8\pi G)^{-1}$
times the part of the Einstein tensor
which is linear in the virtual gravitational field
(that field is defined as the
virtual metric tensor minus the Minkowskian
flat-space metric tensor).
This self-gravitationally
corrected virtual stress-energy tensor
turns out to be non-singular (it actually
vanishes on the light cone), and it Fourier
transforms back to four-momentum space as
the self-gravitationally corrected and
ultraviolet-convergent version of the
insertion Feynman diagram of Fig.~\ref{Fig:VacPol}.
A notable by-product of full Einstein equation
propagation of virtual gravitons, which takes
proper account of their non-linear interactions
with each other, is that the subsequent
perturbative treatment of these
non-linearities is thus
avoided---eliminating the source of the
ultraviolet divergences of the
second-quantized gravity theory itself.

It is worthwhile to point out here that a systematic
general approach to the full theory of self-gravitationally
coupled quantum electrodynamics along the lines discussed
above would involve the usual perturbation expansion in the
purely electromagnetic interaction combined with a stationary
treatment of the gravitational field degrees of freedom in
the Feynman path integration of the total action.  The virtual
stress-energy occurring in the resulting Einstein equation for
the virtual metric could then be simplified by eliminating
terms which do {\em not\/} contribute in an ultraviolet-divergent
fashion to the problem being treated, since any such ``ordinary''
self-gravitational corrections to quantum electrodynamics can be
expected to be completely negligible at attainable energies.%
\footnote{However, see Ref.~\cite{FPVV} for the situation at
energies far above the Planck scale, where the {\em entire\/}
dynamics turns out to reduce to the {\em full\/} effect of
just classical gravitation coupled to classical particle
motion.}  The justification of the key stationary approximation
for the gravitational degrees of freedom in the Feynman path
integration can be argued as follows.  When gravitational effects
are sufficiently weak, gravity theory is essentially linear, and
for linear field theories the stationary approximation to Feynman
path integration in their field variables happens to be exact.
Moreover, the gravitational action contains terms such as
the integral over space--time of the curvature scalar
{\em divided\/} by the factor $16 \pi G$, which become
{\em large\/} compared to $\hbar$ long before
the non-linear corrections to gravity theory, which are
typically suppressed by a factor of $G$ relative to the
linear terms, begin to contribute significantly.  Of
course, the ``classic'' (i.e., Correspondence Principle)
situation which justifies the stationary approximation
to Feynman path integration occurs precisely when the
action is large compared to $\hbar$.  The very
considerable {\em overlap\/} here between this state of
affairs (which always obtains for sufficiently
{\em strong\/} gravity) and the ``effectively linear
theory'' justification for the stationary approximation
(up to the point where the non-linear effects of
gravitation intervene) ensures that the stationary
approximation is {\em uniformly\/} valid for any
strength of gravitational effect which may be present.
It seems plausible that the general approach
envisioned here will essentially parallel that of
Ref.~\cite{ISS}, except that the ``graviton
superpropagator'' of the latter will be replaced by
full Einstein equation propagation of virtual
intermediate gravitons, and perturbative treatment
of non-linear parts of the Einstein equation cannot
arise.  Detailed formal development of this envisioned
systematic general approach to self-gravitationally
coupled quantum electrodynamics is not attempted in
the present paper; that very substantial endeavor is
deferred to a future time.

The organization of this paper
is as follows.  In Section~\ref{Sect:VPMomentum}
the usual four-momentum space treatment of the
insertion diagram of Fig.~\ref{Fig:VacPol}.
is reviewed.  In Section~\ref{Sect:VPConfiguration}
the full details of the analogous treatment in
configuration space--time, where this diagram is
seen to be a virtual stress-energy tensor, are
spelled out (the virtual stress-energy character
of the higher-order versions of this
diagram is demonstrated as well).
In Section~\ref{Sect:SelfGravityCorrection}
this virtual stress-energy is propagated with
the full Einstein equation to obtain its
corresponding virtual metric,
which in turn yields the self-gravitationally
corrected virtual stress-energy.
The consequences for the four-momentum space
diagram itself of this self-gravitational
correction, such as the suppression
of the ultraviolet divergence, the absence
of any contribution to charge
renormalization, and its behaviour at extreme
momentum transfer, are set forth
in Section~\ref{Sect:Discussion}.  There it is
also pointed out that the closely related
second-order electromagnetic vacuum-to-vacuum
amplitude correction diagram of Fig.~\ref{Fig:VacVac}, which
has an extreme (quartic) ultraviolet divergence,
can be shown to vanish identically after the same
self-gravitational correction.  This obviates the
need for the usual fiat injunction that such a
``disconnected'' diagram is to be ``discarded'',
notwithstanding its strongly infinite value.

\section{The ``vacuum polarization'' insertion in momentum space}
\label{Sect:VPMomentum}

In this Section the standard four-momentum
space approach to the ultraviolet-divergent
``vacuum polarization'' radiative correction
insertion diagram shown in Fig.~\ref{Fig:VacPol}
is reviewed.  That diagram consists of a single
photon propagator, one end of which is attached
to one of the two vertices of a
virtual electron--positron pair loop.
This ultraviolet-divergent insertion
may be added between any vertex and
its attached photon line in any
quantum electrodynamics Feynman diagram,
in order to generate one of that
diagram's radiative corrections.  The
second-order ``vacuum polarization''
radiative correction insertion
of Fig.~\ref{Fig:VacPol}
is expressed as \cite{BjorkenDrell}:
\begin{mathletters}
\begin{equation}
\widehat T^{\mu\nu}(q) = {-e^2\over q^2+i\epsilon}
\widehat L^{\mu\nu}(q),
\label{Eq:Insertion}
\end{equation}
where the electron--positron virtual pair loop portion is
\begin{equation}
\widehat L^{\mu\nu}(q)\equiv \,-4\pi i
\,\int{d^4p\over (2\pi)^4}\>{\rm Tr}
\left (\gamma^{\mu}\widehat S_F(p)\gamma^{\nu}
\widehat S_F(p-q)\right ),
\label{Eq:LoopPortion}
\end{equation}
with the spinor propagator $\widehat S_F(p)$ given by
\begin{equation}
\widehat S_F(p) = {\slsh p + m \over p^2 - m^2 +i\epsilon}.
\label{Eq:SpinorPropagator}
\end{equation}
\end{mathletters}%
\setcounter{insertionEq}{\value{equation}}%
{}From Eqs.~(\ref{Eq:LoopPortion})\ and
(\ref{Eq:SpinorPropagator})\ above,
we can see that the loop factor $\widehat
L^{\mu\nu}(q)$ is apparently quadratically
divergent, and that the entire
insertion, $\widehat T^{\mu\nu}(q)$,
is dimensionless, which an insertion
necessarily must be.  If the photon line
which one may attach to the
loop portion of this insertion happens
to be an external one with $q^2 = 0$,
then gauge invariance with respect
to permissible variation of this external
photon's polarization four-vector
implies that we must have
\begin{equation}
\widehat L^{\mu\nu}(q)\,q_{\nu} = 0 \quad\hbox{for $q^2 = 0$}.
\label{Eq:GaugeInvariance}
\end{equation}
It turns out that $\widehat L^{\mu\nu}(q)$
may be reexpressed
in the form \cite{BjorkenDrell2},
\[
\widehat L^{\mu\nu}(q) =
\left(\eta^{\mu\nu}q^2 - q^{\mu}q^{\nu}\right)
\hat h(q^2) + \eta^{\mu\nu}\hat d(q^2),
\]
where we use $\eta^{\mu\nu}$ to denote the
flat-space (Minkowskian)\ metric
tensor.  The term proportional to $\hat h(q^2)$
above automatically satisfies
the gauge-invariance condition of
Eq.~(\ref{Eq:GaugeInvariance}).
In Ref.\ \cite{BjorkenDrell2}, an argument
is made that the object $\hat d(q^2)$ above,
although appearing to be
quadratically divergent, in fact vanishes.
Thus we may write
\begin{mathletters}
\begin{equation}
\widehat L^{\mu\nu}(q) = \left (\eta^{\mu\nu}q^2
- q^{\mu}q^{\nu}\right)
\hat h(q^2)
\label{Eq:Lhat}
\end{equation}
and, from Eq.~(\ref{Eq:Insertion}),
\begin{equation}
\widehat T^{\mu\nu}(q) = {-e^2\over q^2 + i\epsilon}
\left(\eta^{\mu\nu}q^2 - q^{\mu}q^{\nu}\right)
\hat h(q^2),
\label{Eq:That}
\end{equation}
\end{mathletters}%
\setcounter{hatEq}{\value{equation}}%
where $\hat h(q^2)$ is only
logarithmically divergent \cite{BjorkenDrell2}.
We now see
that the insertion $\widehat T^{\mu\nu}(q)$
is symmetric in its two Lorentz
indices $\mu$ and $\nu$, dimensionless,
and always satisfies
\begin{equation}
q_{\mu}\widehat T^{\mu\nu}(q) = 0.
\end{equation}
Thus its configuration space--time
Fourier transform $T^{\mu\nu}(x)$ must
also be symmetric in its two Lorentz indices,
have vanishing four-divergence,
and have the dimensions of $q^4$,
which are those of energy
density.\footnote{We use the
conventional system of units
where $\hbar = c = 1$.}
So $T^{\mu\nu}(x)$ has the attributes
of a stress-energy tensor, albeit a
virtual one---we shall see in
Section~\ref{Sect:VPConfiguration}
that it is complex-valued.

Before entering the detailed calculation of
$T^{\mu\nu}(x)$, which we shall undertake in the
next section, we first need
to show how the logarithmic
divergences of $\widehat L^{\mu\nu}(q)$,
$\widehat T^{\mu\nu}(q)$ and $\hat h(q^2)$
in Eqs.~(\thehatEq) are customarily handled.  One
splits $\hat h(q^2)$ into a
logarithmically divergent part which is
independent of $q^2$, and a $q^2$-dependent
part which is convergent:
\begin{mathletters}
\begin{equation}
\widehat L^{\mu\nu}(q) =
\left(\eta^{\mu\nu}q^2 - q^{\mu}q^{\nu}\right)
\left[ \left.\hat h\right|_{q^2 = 0}
+ \left(\hat h(q^2) - \left.\hat h\right|_{q^2 = 0}\right)
\right]
\label{Eq:Lsplih}
\end{equation}
or
\begin{equation}
\widehat T^{\mu\nu}(q) =
-\,{e^2 \over q^2}\,\left(\eta^{\mu\nu}q^2 -
q^{\mu}q^{\nu}\right)
\left[ \left. \hat h\right|_{q^2 = 0} + \left(\hat h(q^2) -
\left. \hat h\right|_{q^2 = 0}\right)
\right].
\label{Eq:Tsplih}
\end{equation}
\end{mathletters}%
\setcounter{insertionFinalEq}{\value{equation}}%
The logarithmically divergent
$\hat h|_{q^2 = 0}$ is absorbed into charge
renormalization,\footnote{It does seem
questionable to thus imbue an
{\em ultraviolet\/} divergence
with a fundamental r{\^o}le in determining the scale of
the extreme {\em low-energy\/}
scattering strength via such an effective
modification of electronic charge,
even leaving aside the puzzle of how a
diagram which involves but a
single transient virtual pair can manage to
renormalize charge by
effectively polarizing the entire vacuum.} while the
observable $q$-dependent dynamical
effects of this insertion
$\widehat T^{\mu\nu}(q)$ are wholly
attributed to the convergent object
$(\hat h(q^2)
 - \hat h|_{q^2 = 0})$.
As well as being convergent,
$(\hat h(q^2) -  \hat
h|_{q^2 = 0})$ is also proportional to
$q^2$ for sufficiently small $|q^2|$,
i.e., for $|q^2|\ll m^2$,
where $m$ is the electron mass.
For the
purpose of passing to
configuration space--time,
which is where the
self-gravitational correction of
$T^{\mu\nu}(x)$ must be worked out using the Einstein
equation, we rewrite Eq.~(\ref{Eq:Tsplih})
in the form
\begin{equation}
\widehat T^{\mu\nu}(q) =
e^2\left(\eta^{\mu\nu}q^2 -q^{\mu}q^{\nu}\right)
\widehat H(q^2)
- e^2\left(\eta^{\mu\nu} - {q^{\mu}q^{\nu}\over q^2}\right)
\left. \hat h\right|_{q^2 = 0},
\label{Eq:Tsplih2}
\end{equation}
where
\begin{equation}
\widehat H(q^2)\equiv -\,{1\over q^2}\,\left(\hat h(q^2) -
\left.\hat h\right|_{q^2 = 0}\right ).
\label{Eq:Hhatdefined}
\end{equation}
The first term of Eq.~(\ref{Eq:Tsplih2})
is clearly convergent, while the logarithmically
divergent second term will not contribute
to the Fourier transform
$T^{\mu\nu}(x)$ for any $x^{\mu}\ne 0$.
Thus, from Eqs.~(\ref{Eq:Tsplih2}) and (\ref{Eq:Hhatdefined}),
the configuration
space--time Fourier transforms $T^{\mu\nu}(x)$ and $H(x^2)$
are infinity-free for all $x^{\mu}\ne 0$.

\section{The ``vacuum polarization'' insertion in configuration space}
\label{Sect:VPConfiguration}

In Eqs.~(\theinsertionEq)--(\theinsertionFinalEq)
of Section~\ref{Sect:VPMomentum},
the usual four-momentum space treatment of
the ``vacuum polarization'' insertion
was outlined.  The detailed analogous treatment
in configuration space--time is now carried out
in order to obtain the explicit form of
the virtual stress-energy tensor
$T^{\mu\nu}(x)$, whose self-gravitational
correction $T^{\mu\nu}_G (x)$
one subsequently wishes to calculate.

The goal here is to Fourier-transform the
$\hat T^{\mu\nu}(q)$ of Eq.~(\ref{Eq:Insertion}) to
configuration space--time, but the initial effort
will concentrate on the Fourier
transform of just its
loop factor $\hat L^{\mu\nu}(q)$.  Using Eq.~(\ref{Eq:LoopPortion})
we obtain
\begin{eqnarray}
L^{\mu\nu}(x)\!\! &=& \!\! -4\pi i\;{\rm Tr}\left [\int{d^4q\over (2\pi)^4}\,
e^{-iq\cdot x}\int{d^4p\over (2\pi)^4}\> \gamma^{\mu}
\widehat S_F(p)\gamma^{\nu}\widehat S_F(p - q)\right ]\nonumber\\
&=& \!\! -4\pi i\;{\rm Tr}\left [\int{d^4q\over (2\pi)^4}\> e^{-iq\cdot
x}\int{d^4p\over (2\pi)^4}\> \gamma^{\mu}\;\right .\nonumber\\
& &\left .\qquad\qquad\qquad\times\;\int d^4x_1\, e^{ip\cdot x_1}
S_F(x_1)\gamma^{\nu}\int d^4x_2\,
e^{i(p - q)\cdot x_2}S_F(x_2)\right ]\nonumber\\
&=& \!\! -4\pi i\;{\rm Tr}\left [\int d^4x_1\,\int d^4x_2\,\gamma^{\mu}
S_F(x_1)\gamma^{\nu}S_F(x_2)\,\delta^{(4)}(x + x_2)\,\delta^{(4)}
(x_1 + x_2)\right ]\nonumber\\
&=&  \!\!-4\pi i\;{\rm Tr}\left [\gamma^{\mu}S_F(x)\gamma^{\nu}S_F(-x)
\right ],
\label{Eq:LoopFactorx}
\end{eqnarray}
where, using Eq.~(\ref{Eq:SpinorPropagator}),
\begin{eqnarray}
S_F(x)\!\! &=&\!\! \int{d^4p\over (2\pi)^4}\>
e^{-ip\cdot x}\widehat S_F(p)\nonumber\\
&=& \!\!\int{d^4p\over (2\pi)^4}\> e^{-ip\cdot x}
\left(\frac{\slsh p +m}{p^2
- m^2 +i\epsilon}\right)\nonumber\\
&=& \!\!(i\slsh\partial + m)\int{d^4p\over (2\pi)^4}\>
{e^{-ip\cdot x}\over p^2
- m^2 +i\epsilon}\nonumber\\
&=& \!\!(i\slsh\partial + m)\Delta_F(x^2).
\label{Eq:SFx}
\end{eqnarray}
It is convenient to have a
manifestly scalar representation of the
function $\Delta_F(x^2)$.
With some of the details of the variable
changes and intermediate
Gaussian integration steps left to the reader,
its derivation is as follows:
\begin{eqnarray}
\Delta_F(x^2) \!\!&=&\!\!
\int{d^4p\over (2\pi)^4}\, {e^{-ip\cdot x}\over p^2
- m^2 +i\epsilon}\nonumber\\
&=& \! \! -i\, \int_0^{\infty} ds\,\int{d^4p\over (2\pi)^4}\,
e^{-ip\cdot x}
\> e^{i(p^2 - m^2 +i\epsilon )s}\nonumber\\
&=& \!\!-\, {1\over 16\pi^2}\,\int_0^{\infty}{ds\over s^2}
\, e^{-ix^2/(4s)}\> e^{-i(m^2 - i\epsilon )s}\nonumber\\
&=& \!\!-\, {m^2\over 4\pi^2}\,\int_0^{\infty}d\kappa
\, e^{-i\kappa m^2x^2}\> e^{-i(1 - i\epsilon )/(4\kappa)}.
\label{Eq:DeltaFRaw}
\end{eqnarray}
At $x^2 = 0$, the above integral
representation of $\Delta_F(x^2)$ diverges,
so a $\delta$-parametrized
damping factor is inserted:
\begin{equation}
\Delta_F(x^2) = \lim_{\epsilon ,\delta\to 0+}\>
-{m^2\over 4\pi^2}\,\int_0^{\infty}d\kappa\,
e^{-i\kappa (m^2x^2-i\delta )}
\> e^{-i(1 - i\epsilon )/(4\kappa )}.
\label{Eq:DeltaF}
\end{equation}
Near the light cone, $\Delta_F(x^2)$ has the asymptotic form
\begin{mathletters}
\begin{equation}
\Delta_F(x^2)\sim \lim_{\delta\to 0+}\>{i\over 4\pi^2}
\left(\frac{1}{x^2 - i
\delta}\right)\quad\hbox{as $x^2\to 0$}
\label{Eq:DeltaFLightConeA}
\end{equation}
or
\begin{equation}
\Delta_F(x^2)\sim {i\over 4\pi^2}\,{\cal P}{1\over x^2}
- {1\over 4\pi}\,
\delta (x^2)\quad\hbox{as $x^2\to 0$}.
\label{Eq:DeltaFLightConeB}
\end{equation}
\end{mathletters}%
With $\Delta_F(x^2)$ now in
the desired representation given by Eq.~(\ref{Eq:DeltaF}),
we return our attention to the loop tensor $L^{\mu\nu}(x)$
of Eq.~(\ref{Eq:LoopFactorx}),
insert Eq.~(\ref{Eq:SFx})
into it, and carry out the
trace using the trace theorems of Ref.~\cite{BjorkenDrell4}:
\begin{eqnarray}
L^{\mu\nu}(x)\!\! &=&\!\! -4\pi i\;{\rm Tr}{
\left[\gamma^{\mu}S_F(x)\gamma^{\nu}
S_F(-x)\right ]}\nonumber\\
&=& \!\! -4\pi i\;{\rm Tr}\left [\gamma^{\mu}\left ((i\slsh\partial + m)
\Delta_F(x^2)\right )\gamma^{\nu}\left ((-i\slsh\partial + m)
\Delta_F(x^2)\right )\right ]\nonumber\\
&=& \!\!-16\pi i\,\left[2(\partial^{\mu}\Delta_F)
(\partial^{\nu}\Delta_F)-
\eta^{\mu\nu}(\partial_{\alpha}
\Delta_F)(\partial^{\alpha}\Delta_F) +
\eta^{\mu\nu}m^2(\Delta_F)^2\right ],
\label{Eq:LoopFactorTraced}
\end{eqnarray}
where $\eta^{\mu\nu}$ is the
flat-space (Minkowskian) metric tensor.  We
see that $L^{\mu\nu}(x)$
is symmetric in its two Lorentz indices.  If we
take the four-divergence of $L^{\mu\nu}(x)$, we obtain
\begin{eqnarray}
\partial_{\nu}L^{\mu\nu}(x) \!\! &=&\!\! -16\pi i\,
\left[2(\partial^{\mu}
\Delta_F)(\partial_{\nu}\partial^{\nu}\Delta_F)
+ 2m^2(\partial^{\mu}
\Delta_F)\Delta_F\right ]\nonumber\\
&=& \!\! -16\pi i\,\left [2(\partial^{\mu}\Delta_F)
\left((\partial_{\nu}
\partial^{\nu} + m^2)\Delta_F\right )\right ]\nonumber\\
&=&  \!\!-16\pi i\,
\left[2\left(\partial^{\mu}\Delta_F(x^2)\right)
\left(
-\delta^{(4)}(x)\right )\right ]\nonumber\\
&=& \!\! -16\pi i\,\left [-4 x^{\mu}\Delta_F' (x^2)\,
\delta^{(4)}(x)\right ],
\label{Eq:DivL}
\end{eqnarray}
where we have used the fact that the
basic definition of $\Delta_F$ (see
Eq.~(\ref{Eq:SFx})) implies that it is a
Green function for the Klein--Gordon
equation:
$(\partial_{\nu}\partial^{\nu} + m^2)\Delta_F
= -\delta^{(4)}(x)$.
We see from Eq.~(\ref{Eq:DivL}) that
\begin{equation}
\partial_{\nu}L^{\mu\nu}(x) = 0 \quad\hbox{if $x \ne 0$}.
\label{Eq:DivLNull}
\end{equation}
Also, from Eq.~(\ref{Eq:DeltaF}) we obtain
\begin{equation}
\Delta_F' (x^2) =
{i\left (m^2\right )^2\over 4\pi^2}\,\int_0^{\infty}\kappa\, d\kappa\,
e^{-i\kappa (m^2x^2-i\delta )}\> e^{-i(1-i\epsilon )/(4\kappa )}.
\label{Eq:DeltaFp}
\end{equation}
We thus note that the object $x^{\mu}\Delta_F' (x^2)$ is of odd
parity, and, if we hold off taking $\delta\to 0$
in Eq.~(\ref{Eq:DeltaFp}), it is even
equal to zero when $x = 0$.
Thus we are at least close to an argument
that $x^{\mu}\Delta_F' (x^2)\,\delta^{(4)}(x)$ actually
vanishes at $x = 0$, so that, with Eq.~(\ref{Eq:DivLNull}),
we obtain $\partial_{\nu}
L^{\mu\nu}(x) = 0$ everywhere.
Of course, because our insertion must be
electromagnetically gauge-invariant,
we do need to require that $L^{\mu
\nu}(x)$ indeed has such a
vanishing four-divergence.  We now wish to
recast $L^{\mu\nu}(x)$ into a simple
form whose vanishing four-divergence
is manifest.  Using the fact that
$\partial^{\mu}\Delta_F(x^2)
= 2x^{\mu}\Delta_F' (x^2)$,
we first rewrite Eq.~(\ref{Eq:LoopFactorTraced}) as
\begin{eqnarray}
L^{\mu\nu}(x)\! \! & = & \! \!
-16\pi i\,\left\{x^{\mu}x^{\nu}\left(8
(\Delta_F' (x^2))^2\right) \right. \nonumber \\
& & \!\! \left.\quad\quad\quad\quad {}
-\eta^{\mu\nu}\left[x^28(\Delta_F'
(x^2))^2
- \left (x^24(\Delta_F' (x^2))^2
+ m^2(\Delta_F(x^2))^2\right )\right]\right\}.
\label{Eq:LoopTensor}
\end{eqnarray}
Now it may readily be demonstrated that a tensor of the form
\begin{mathletters}
\begin{equation}
A^{\mu\nu}(x) = x^{\mu}x^{\nu}f(x^2) - \eta^{\mu\nu}\left (x^2f(x^2) +
d(x^2)\right)
\label{Eq:ATensor}
\end{equation}
has vanishing four-divergence if and only if
\begin{equation}
d'(x^2) = {3\over 2}\, f(x^2),
\label{Eq:ATensorConditionRaw}
\end{equation}
that is, if
\begin{equation}
d(x^2) = {3\over 2}\,\int_{{x_1}^2}^{x^2}d\lambda\, f(\lambda).
\label{Eq:ATensorCondition}
\end{equation}
\end{mathletters}%
\setcounter{vanishingFlatDivergenceEq}{\value{equation}}%
We thus see that the
divergenceless nature of $A^{\mu\nu}(x)$ effectively
means that it depends only on a
single scalar function.  Can we therefore
make
$A^{\mu\nu}(x)$
{\em manifestly\/}
divergenceless
by writing it
in terms
of a single scalar function?
To this end we look at the manifestly
divergenceless form:
\begin{mathletters}
\begin{equation}
B^{\mu\nu}(x) = \left (\partial^{\mu}\partial^{\nu}
- \eta^{\mu\nu}
\partial_{\alpha}\partial^{\alpha}\right )h(x^2),
\label{Eq:BTensorRaw}
\end{equation}
which yields
\begin{equation}
B^{\mu\nu}(x) = x^{\mu}x^{\nu}4h''(x^2) - \eta^{\mu\nu}
\left (x^24h''(x^2) + 6h'(x^2)\right ).
\label{Eq:BTensor}
\end{equation}
\end{mathletters}%
Comparison of Eq.~(\ref{Eq:BTensor})
with Eqs.~(\thevanishingFlatDivergenceEq) shows that
$B^{\mu\nu}(x) =
A^{\mu\nu}(x)$ if
\[
h''(x^2) = {1\over 4}\, f(x^2)
\]
and
\[
h'(x^2) = {1\over 6}\, d(x^2) = {1\over 4}\,
\int_{{x_1}^2}^{x^2}d\lambda\, f(\lambda),
\]
which amount to the same thing.  So we simply require that
\begin{equation}
h(x^2) = {1\over 4}\,\int_{{x_0}^2}^{x^2}d\lambda\,
\int_{{x_1}^2}^{\lambda}d\lambda '\, f(\lambda ').
\label{Eq:hrequirement}
\end{equation}
To reexpress Eq.~(\ref{Eq:LoopTensor}) in the
manifestly divergenceless format of
Eq.~(\ref{Eq:BTensorRaw}), the $f(x^2)$ we need is
\begin{equation}
f(x^2) = -128\pi i\,\left (\Delta_F'(x^2)\right )^2,
\label{Eq:fneed}
\end{equation}
and, as we wish to arrange that
$h(x^2)$ vanishes as $|x^2|\to \infty$
(this will later permit us to
obtain $H(x^2)$ having the same property),
we take both ${x_0}^2$ and
${x_1}^2$ in Eq.~(\ref{Eq:hrequirement}) to be $-\infty$.  Using
Eq.~(\ref{Eq:DeltaFp}), we can express
Eq.~(\ref{Eq:fneed}) in the form
\begin{equation}
f(x^2) =
{8i\left (m^2\right )^4\over \pi^3}\,\int_0^{\infty}d\kappa_1\,
\int_0^{\infty}d\kappa_2\,\kappa_1\kappa_2\,
e^{-i(\kappa_1 + \kappa_2)(m^2x^2 - i\delta )}\>
e^{-i(1-i\epsilon )(\kappa_1 + \kappa_2)/(4\kappa_1\kappa_2)}.
\label{Eq:fexplicitRaw}
\end{equation}
It is now convenient to change variables in
Eq.~(\ref{Eq:fexplicitRaw}) to $\kappa\equiv
\kappa_1 + \kappa_2$ and $z\equiv
\kappa_1/(\kappa_1 +\kappa_2)$.  Thus,
$\kappa_1 = z\kappa$,
$\kappa_2 = (1 -z)\kappa $, and $d\kappa_1\, d
\kappa_2 = dz\, \kappa\, d\kappa$.
Equation (\ref{Eq:fexplicitRaw}) becomes
\begin{mathletters}
\begin{equation}
f(x^2) =
{8i\left (m^2\right )^4\over \pi^3}\,
\int_0^{\infty}\, d\kappa\, \kappa^3\,
e^{-i\kappa (m^2x^2 - i\delta )}\,Z_{\epsilon}(\kappa ),
\label{Eq:fexplicit}
\end{equation}
where
\begin{equation}
Z_{\epsilon}(\kappa )\equiv \int_0^1 dz\, z(1 - z)\,
e^{-i(1 - i\epsilon )/(4\kappa z(1 - z))}.
\label{Eq:Zdef}
\end{equation}
\end{mathletters}%
\setcounter{fofZEq}{\value{equation}}%
Using the Taylor expansion and the
stationary phase approximation, one readily
obtains the asymptotic behaviours of $Z_{\epsilon}(\kappa )$:
\begin{mathletters}
\begin{eqnarray}
Z_{\epsilon}(\kappa )
\!\!&\sim&\!\! {1\over 6}\,
\left(1 - {3i\over 2\kappa}
\right)\quad\hbox{as $\kappa\to\infty$},
\label{Eq:Zinfinity}\\
Z_{\epsilon}(\kappa )
\!\! &\sim& \!\!
{e^{-i\pi /4}\over 8}\,(\pi\kappa)^{1\over 2}\,
e^{-i(1 - i\epsilon )/\kappa}\quad\hbox{as $\kappa\to 0$}.
\label{Eq:Zzero}
\end{eqnarray}
\end{mathletters}%
\setcounter{ZsimEq}{\value{equation}}%
While Eqs.~(\theZsimEq)
tell us that $Z_{\epsilon}(\kappa )$
approaches a constant
as $\kappa\to\infty$, it has an
oscillatory essential singularity
as $\kappa\to 0$ that can
effectively cut off singularities which other
functions, integrated over
$\kappa $ together with $Z_{\epsilon}(\kappa )$,
may have in this limit.
Examples of such $\kappa\to 0$ singularities occur
when we use Eqs.~(\ref{Eq:hrequirement})
and (\thefofZEq)
to evaluate our $h(x^2)$.  In particular,
we have
\begin{mathletters}
\begin{eqnarray}
\int_{-\infty}^{x^2}d\lambda\, e^{-i\kappa m^2\lambda}
\!\! &= &\!\!
\lim_{\Lambda\to +\infty}\int_{-\Lambda}^{x^2}d\lambda\,
e^{-i\kappa m^2\lambda}\nonumber\\
&=& \!\!{1\over m^2}\,\left (\pi\delta (\kappa )
+ i e^{-i\kappa m^2x^2}\,
{\cal P}{1\over\kappa}\right ),
\label{Eq:IntegralIdentity}
\end{eqnarray}
and
\begin{equation}
\lim_{x^2\to\pm\infty}\left (i e^{-i\kappa m^2x^2}\,{\cal P}
{1\over\kappa}\right ) = \pm\pi\delta (\kappa ).
\label{Eq:PIdentity}
\end{equation}
\end{mathletters}%
The presence of the oscillatory
essential singularity in $Z_{\epsilon}
(\kappa )$ as $\kappa\to 0$ always
permits us to drop singular niceties
such as the $\pi\delta (\kappa )$ and
the principal-value notation in
Eq.~(\ref{Eq:IntegralIdentity}) when we carry
out this sort of integration under the integral
sign in representations of the
type given by Eq.~(\ref{Eq:fexplicit}).
Thus, for
the desired $h(x^2)$
associated with the $f(x^2)$ of
Eqs.~(\ref{Eq:fneed}) and (\thefofZEq),
we obtain
\begin{eqnarray}
h(x^2) \!\! &=& \!\!
{1\over 4}\,\int_{-\infty}^{x^2}d\lambda\,
\int_{-\infty}^{\lambda}d\lambda '\, f(\lambda ')\nonumber\\
&=& \!\! {-2i\left (m^2\right )^2\over \pi^3}\,\int_0^{\infty}\,\kappa\,
d\kappa\, e^{-i\kappa (m^2x^2 - i\delta )}\,Z_{\epsilon}(\kappa ).
\label{Eq:hexplicit}
\end{eqnarray}
Equation (\ref{Eq:hexplicit})
provides an explicit expression for the $h(x^2)$ which permits
us to write the loop tensor $L^{\mu\nu}(x)$ of
Eq.~(\ref{Eq:LoopTensor}) in the manifestly
divergenceless format of Eq.~(\ref{Eq:BTensorRaw})
\begin{equation}
L^{\mu\nu}(x) = \left (\partial^{\mu}\partial^{\nu} - \eta^{\mu\nu}
\partial_{\alpha}\partial^{\alpha}\right )h(x^2).
\label{Eq:LoopTensorofh}
\end{equation}
Equation (\ref{Eq:LoopTensorofh}) is, of course,
the space--time version of Eq.~(\ref{Eq:Lhat}).
We also wish to explicitly
obtain the virtual stress-energy tensor $T^{\mu\nu}(x)$
itself in the similar manifestly divergenceless form
given by
\begin{mathletters}
\begin{equation}
T^{\mu\nu}(x) = e^2\,\left(\partial^{\mu}\partial^{\nu} - \eta^{\mu\nu}
\partial_{\alpha}\partial^{\alpha}\right )H(x^2),
\label{Eq:TmunuFlat}
\end{equation}
where, from Eq.~(\ref{Eq:Insertion}) and
Eq.~(\ref{Eq:LoopTensorofh}), we must have that
\begin{equation}
\partial_{\alpha}\partial^{\alpha} H(x^2) = h(x^2).
\label{Eq:Heq}
\end{equation}
\end{mathletters}%
\setcounter{FlatTmunuEq}{\value{equation}}%
We note here that since
$H(x^2)$ is the four-dimensional
Fourier transform of the finite function
$\widehat H(q^2)$ of Eq.~(\ref{Eq:Hhatdefined}), which
is well-behaved as $q^2\to 0$,
we can infer that $H(x^2)\to 0$ as $|x^2|
\to \infty$.
To explicitly obtain the stress-energy tensor (\ref{Eq:TmunuFlat}),
we must solve the d'Alembertian relation of
Eq.~(\ref{Eq:Heq}) for the function
$H(x^2)$, given our $h(x^2)$ of Eq.~(\ref{Eq:hexplicit}).
Thus we must solve the equation
\[
\partial_{\alpha}\partial^{\alpha}H(x^2) =
4\left ({d\over dx^2}\right )^2
\left (x^2\, H(x^2)\right ) = h(x^2),
\]
for which we readily obtain a solution in
accord with the requirement noted
above that $H(x^2)$ vanish as $|x^2|\to \infty$,
\begin{mathletters}
\begin{eqnarray}
H(x^2) \!\! &=& \!\!
{1\over 4 x^2}\,\int_{-\infty}^{x^2}d\lambda\,
\int_{-\infty}^{\lambda}
d\lambda '\, h(\lambda ')
\label{Eq:HsolutionImplicit}\\
&=& \!\! {i\over 2 \pi^3 x^2}\,\int_0^{\infty}\, {d\kappa\over \kappa}\,
e^{-i\kappa (m^2x^2 - i\delta )}\,Z_{\epsilon}(\kappa ).
\label{Eq:HsolutionExplicit}
\end{eqnarray}
\end{mathletters}%
\setcounter{HofxEq}{\value{equation}}%
Equation (\ref{Eq:HsolutionExplicit})
explicitly provides the desired $H(x^2)$ for the representation
of $T^{\mu\nu}(x)$, which is given by Eq.~(\ref{Eq:TmunuFlat}).
These equations make it clear that $T^{\mu\nu}(x)$ has the
dimensions of stress-energy as well as the stress-energy
tensor properties of index symmetry and divergencelessness.
$T^{\mu\nu}(x)$ is, however, a {\em virtual\/}
stress-energy---its complex-valued nature follows
from that of $H(x^2)$.

With the desired explicit representations of
$h(x^2)$ and $H(x^2)$ in hand
in Eqs.~(\ref{Eq:hexplicit}) and
(\theHofxEq), it is useful to
obtain their asymptotic behaviours.\footnote{Standard
asymptotic techniques are applied here,
including the stationary-phase
approximation, the saddle-point
approximation, and taking subexpression
limits after changes of variable.}
The results obtained for $h(x^2)$ are
\begin{mathletters}
\begin{eqnarray}
h(x^2) \!\! &\sim & \!\!
{-m\over 4\pi^2 (x^2)^{3\over 2}}\, e^{-2im\sqrt{x^2}}\quad
\hbox{as $x^2\to +\infty$},
\label{Eq:hPlusInfinity}\\
h(x^2) \!\! &\sim& \!\!
{im\over 4\pi^2 (-x^2)^{3\over 2}}\, e^{-2m\sqrt{-x^2}}\quad
\hbox{as $x^2\to -\infty$},
\label{Eq:hMinusInfinity}\\
h(x^2) \! &\sim & \! {i\over 3\pi^3 (x^2 - i\delta )^2}\quad
\hbox{as $x^2\to 0$},
\label{Eq:hLightCone}
\end{eqnarray}
\end{mathletters}%
\setcounter{hsimEq}{\value{equation}}%
and those for $H(x^2)$ are:
\begin{mathletters}
\begin{eqnarray}
H(x^2) \!\! &\sim& \!\!
{1\over 16\pi^2 m(x^2)^{3\over 2}}\, e^{-2im\sqrt{x^2}}\quad
\hbox{as $x^2\to +\infty$},
\label{Eq:HPlusInfinity}\\
H(x^2)\!\! &\sim& \!\!
{-i\over 16\pi^2 m(-x^2)^{3\over 2}}\, e^{-2m\sqrt{-x^2}}\quad
\hbox{as $x^2\to -\infty$},
\label{Eq:HMinusInfinity}\\
H(x^2)\!\! &\sim& \!\!
{-i\ln (m^2x^2 - i\delta )\over 12\pi^3 x^2}\quad
\hbox{as $x^2\to 0$}.
\label{Eq:HLightCone}
\end{eqnarray}
\end{mathletters}%
\setcounter{HsimEq}{\value{equation}}%

Equation~(\ref{Eq:hLightCone}) shows, as was
pointed out in Section~\ref{Sect:VPMomentum},
that the Fourier
transform to four-momentum space of $h(x^2)$,
namely $\hat h(q^2)$, diverges
logarithmically. This is due to the
non-integrable light-cone singularity (as $x^2\to 0$).
However, Eq.~(\ref{Eq:HLightCone})
shows that the light-cone singularity of $H(x^2)$ is
integrable over space--time,
so that its Fourier transform $\widehat H(q^2)$ is
finite.  This Fourier transform
$\widehat H(q^2)$ may readily be obtained by
using our representation for
$H(x^2)$ in Eq.~(\ref{Eq:HsolutionExplicit}).
The integration over
configuration space--time becomes essentially a
straightforward Gaussian
integration exercise (it is helpful to
first introduce an additional auxiliary
integration, which raises the leading
factor of $(x^2)^{-1}$ into the exponential
as well---this auxiliary integral is
then straightforwardly evaluated
after the Gaussian integrations over
space--time have been carried out).
Once the space--time integrations have
been carried out, the $\kappa$-integration
of Eq.~(\ref{Eq:HsolutionExplicit}) may also
readily be evaluated, provided one keeps the
auxiliary integration over $z$ which occurs
in the definition of
$Z_{\epsilon}(\kappa )$---see Eq.~(\ref{Eq:Zdef}).
One can as well carry out the
final $z$-integration analytically,
but the properties of the result are
more transparent if one eschews
this step in favour of keeping the
$z$-integral form.  Although $\hat h(q^2)$
diverges logarithmically, as we
have noted, one can as well
take the representation of Eq.~(\ref{Eq:hexplicit}) for
$h(x^2)$ and at least begin the
analogous procedure.  The Gaussian
space--time integrations are
readily carried out, but the subsequent
$\kappa$-integration is then seen to
diverge logarithmically at its upper
limit.  At this stage,
before the $\kappa$-integration is carried out,
one can instead write down the representation for
$(\hat h(q^2) - \hat h|_{q^2 = 0})$.
For this object the $\kappa$-integration
in fact converges---indeed one
recognizes that the resulting expression is
closely related to that obtained for $\widehat H(q^2)$:
\begin{mathletters}
\begin{eqnarray}
\left (\hat h(q^2) - \left.\hat h\right|_{q^2 = 0}\right )
\!\! &=& \!\! -q^2\widehat H(q^2)
\label{Eq:HDiffCheck}\\
&=&\!\! -\,{2\over\pi}\,\int_0^1dz\,z(1-z)\ln\left (1 - {q^2z(1-z)\over m^2
(1 - i\epsilon )}\right ).
\label{Eq:hDiffExplicit}
\end{eqnarray}
\end{mathletters}%
Equation (\ref{Eq:HDiffCheck})
checks with Eq.~(\ref{Eq:Hhatdefined}) of
Section~\ref{Sect:Introduction}, and Eq.~(\ref{Eq:hDiffExplicit})
checks against
standard results for the ``vacuum polarization''
Feynman diagram \cite{BjorkenDrell5} upon
substitution into Eq.~(\theinsertionFinalEq) or
Eq.~(\ref{Eq:Tsplih2}).  Thus the Fourier transform
exercise that we have sketched above in
words validates the correctness of our representations
of $h(x^2)$ and $H(x^2)$ given by Eqs.~(\ref{Eq:hexplicit}) and
(\theHofxEq), and also
the properties of these functions
which follow from those representations
(e.g.\ Eqs.~(\thehsimEq) and (\theHsimEq)).

The basic second-order virtual stress-energy $T^{\mu\nu}(x)$
that is being discussed here, as well as versions of it which are
corrected to higher order in the electromagnetic coupling strength
$e$, may both be compactly characterized in the language of
second-quantized fields,
\begin{mathletters}
\begin{eqnarray}
T^{\mu\nu}(x)
\!\! &=& \!\! \langle 0|\, T(\, A^{\mu}(x)\, J^{\nu}(0)\, )\, |0\rangle
\label{Eq:QFTmunuAJ}\\
&=&\!\! \int d^4x'\, D_F((x - x')^2)\> \langle 0|\, T(\, J^{\mu}(x')\,
J^{\nu}(0)\, )\, |0\rangle ,
\label{Eq:QFTmunuJJ}
\end{eqnarray}
\end{mathletters}%
\setcounter{QFTmunuEq}{\value{equation}}%
where $|0\rangle $ is the vacuum state, $T$ is the Dyson
time-ordering symbol, $D_F(x^2)$ is the
Lorentz-gauge photon propagator (this object is equal to
the zero-mass limit of $-\Delta_F(x^2)$ and satisfies
$\partial_{\alpha} \partial^{\alpha} D_F(x^2) =
\delta^{(4)}(x)$), $A^{\mu}(x)$ is, ideally, the
second-quantized Heisenberg-picture electromagnetic
vector potential in Lorentz gauge, and $J^{\mu}(x)$
is, ideally, its corresponding second-quantized
Heisenberg-picture source current density.  Although
it isn't known how to practically construct the full
Heisenberg-picture source current density $J^{\mu}(x)$,
we {\em can\/} at least suppose we have obtained it through
some finite perturbative order $n$ in $e$.  Up to this
point we have, of course, been dealing with the case
$n =1$, since the lowest possible order perturbative
approximation to $J^{\mu}(x)$ in quantum electrodynamics
is an object that is simply proportional to $e$.

Just as in the $n = 1$ case, it is clear from
Eq.~(\ref{Eq:QFTmunuJJ}) that $T^{\mu\nu}(x)$ is
always simply related to its generalized ``loop
portion'', which is
\begin{equation}
L^{\mu\nu}(x) \equiv \, e^{-2} \langle 0|\, T(\, J^{\mu}(x)\,
J^{\nu}(0)\, )\, |0\rangle .
\label{Eq:QFLmunuDef}
\end{equation}
{}From Ref.~\cite{BjorkenDrell6} we learn that through use of
spectral decomposition of the above expression for $L^{\mu\nu}(x)$
(i.e., insertion of a complete set of states, that all have definite
four-momenta, between $J^{\mu}(x)$ and $J^{\nu}(0)$, followed by
application onto these states of the space--time translation
operator which transforms $J^{\mu}(x)$ to $J^{\mu}(0)$), combined
with the $TCP$ transformation properties of the current density
$J^{\mu}(x)$, it can be shown that $L^{\mu\nu}(x)$ is symmetric
in its indices $\mu$ and $\nu$.  Conservation of the current
density $J^{\mu}(x)$ implies that $L^{\mu\nu}(x)$ is also
divergenceless.  The upshot is that this generalized $L^{\mu\nu}(x)$
of Eq.~(\ref{Eq:QFLmunuDef}) can be expressed in the same general
form as we have in Eq.~(\ref{Eq:LoopTensorofh}) for the $n = 1$
case:
\begin{equation}
L^{\mu\nu}(x) = \left (\partial^{\mu}\partial^{\nu} - \eta^{\mu\nu}
\partial_{\alpha}\partial^{\alpha}\right )h(x^2)\, .
\label{Eq:QFLmunuofh}
\end{equation}
{}From this and Eqs.~(\ref{Eq:QFTmunuJJ}) and (\ref{Eq:QFLmunuDef})
above we obtain as well the generalization to higher order of the
forms of Eqs.~(\theFlatTmunuEq):
\begin{mathletters}
\begin{equation}
T^{\mu\nu}(x) = \, e^2\,\left(\partial^{\mu}\partial^{\nu} - \eta^{\mu\nu}
\partial_{\alpha}\partial^{\alpha}\right )H(x^2)
\label{Eq:QFTmunuofH}
\end{equation}
where
\begin{equation}
\partial_{\alpha}\partial^{\alpha} H(x^2) = h(x^2)\, .
\label{Eq:QFHeq}
\end{equation}
\end{mathletters}%
Unlike in the case of $n = 1$, we don't, of course, have explicit
representations for the $h(x^2)$ and $H(x^2)$ that correspond to
general order $n$, nor do we know the asymptotic
behaviours of these functions.  However, it is quite plausible
that general theorems exist that can at least provide those
asymptotic behaviours, which would likely be sufficient for
obtaining the main features of the self-gravitational corrections
of these higher order cases by means of the approach of the next
Section.  The self-gravitational correction of $T^{\mu\nu}(x)$
corrected to higher electromagnetic orders would therefore
seem to be an interesting topic for future investigation.

It is worth pointing out here that the form given for
$T^{\mu\nu}(x)$ in Eq.~(\ref{Eq:QFTmunuAJ}), notwithstanding
its delocalized and complex-valued ``virtual quantum'' nature,
also bears a strong resemblance to the Lagrangian density's
electromagnetic {\em interaction} term that contains the
coupling of the vector potential to the current density.
In view of the vacuum expectation value that also is a
feature of Eq.~(\ref{Eq:QFTmunuAJ}), it would thus seem
appropriate to refer to $T^{\mu\nu}(x)$ as the vacuum virtual
electromagnetic interaction stress-energy.

\section{The self-gravitationally corrected virtual stress-energy tensor}
\label{Sect:SelfGravityCorrection}

The vacuum virtual electromagnetic interaction stress-energy
tensor $T^{\mu\nu}(x)$ for $n=1$ may be expected, because of
its non-integrable singularity on the light cone, to give rise to
{\em ultrastrong} virtual self-gravitational effects.  Under
such circumstances, approximate treatments of the propagation
of resulting intermediate virtual gravitons which ignore the
non-linear feature of their having interactions {\em with
each other\/} can only be regarded with great misgiving.
Thus this propagation will be effected here by means
of the full Einstein equation, so that it is described
in terms of a virtual metric.
A notable by-product of this approach
is that subsequent approximate
perturbative treatment of these non-linear
graviton-graviton interactions is thus
avoided, which eliminates the source of the
ultraviolet divergences which occur in the
usual approaches to
second-quantized gravity theory itself.
As the virtual gravitational
source $T^{\mu\nu}(x)$ is dependent on
just $H(x^2)$, a single
function of only the Lorentz
invariant variable $x^2\equiv \eta_{\mu\nu} x^{\mu}x^{\nu}$,
one can expect the corresponding virtual metric to have
an analogously high degree of symmetry.
Thus a conformally flat virtual metric
ansatz whose departure from flatness
depends on a single function of $x^2$
is the natural one to try:
\begin{equation}
g_{\mu\nu} = [B(x^2)]^{-2}\eta_{\mu\nu}.
\label{Eq:Metric}
\end{equation}
It is now straightforward, if somewhat tedious,
to calculate the virtual Einstein tensor for this
$g_{\mu\nu}$ (the sign conventions used here for curvature
tensors are those of Weinberg \cite{Weinberg3}):
\begin{equation}
G^{\mu\nu} = -8B^3B''x^{\mu}x^{\nu} - g^{\mu\nu}
\left [-8B^3B''x^2B^{-2} +
12 BB'\left (BB'x^2B^{-2} -1\right )\right ],
\label{Eq:EinsteinTnsr}
\end{equation}
where the
$x^{\mu}\equiv \eta^{\mu\nu}(\partial x^2/\partial x^{\nu})/2$
are a direct generalization of the Car\-te\-sian coordinates
of Min\-kow\-ski space--time to our conformally flat case
(recall that $x^2\equiv \eta_{\mu\nu}x^{\mu}x^{\nu}$).
The three manifestly invariant objects $g_{\mu\nu}G^{\mu\nu}$,
$G_{\mu\nu}G^{\mu\nu}$, and $\det(G_{\mu\nu})/\det(g_{\mu\nu})$
which can be formed from $G^{\mu\nu}$ and $g_{\mu\nu}$ may be
worked out in terms of $x^2$, $B$, $B'$, and $B''$.  From these
three expressions it can be concluded that the three
algebraically simpler objects $BB'$, $x^2B^{-2}$, and $B^3B''$
are also invariants (but we note that neither $B$ nor $x^2$
themselves are invariants).  Taking note of these three simple
invariants and of the contravariant tensor character of our
$G^{\mu\nu}$, whose form as given in Eq.~(\ref{Eq:EinsteinTnsr})
involves these invariants as coefficients, we conclude that
$x^{\mu}x^{\nu}$ is also a contravariant tensor (whose
contraction $g_{\mu\nu}x^{\mu}x^{\nu}$ evaluates to one of our
three simple invariants, namely $x^2B^{-2}$).

The flat-space virtual stress-energy tensor
(\ref{Eq:TmunuFlat}) now needs to be generalized
to the curved space--time geometry specified above.
The standard procedure is the ``minimal coupling''
prescription, according to which one must replace
partial derivatives with covariant derivatives,
and the Minkowski metric with the curved space--time
metric.  In the present case, however,
the covariant divergence of the minimal coupling
expression does not vanish, but is proportional
to the four-gradient of $H$ contracted with the
Ricci curvature tensor.
Thus the minimal coupling recipe cannot be applied to
the Fourier transform of the ``vacuum polarization'' diagram.%
\footnote{The physical reason for the inapplicability of
the minimal-coupling prescription in this instance is
that our $T^{\mu\nu}$ describes a (virtual) {\em extended\/}
system, as is apparent from the presence of the virtual
electron--positron loop in the Feynman diagram, as
well as the explicitly delocalized character of the
representations of $T^{\mu\nu}(x)$ in Eqs.~(\theQFTmunuEq).
Such extended systems are subject to {\em tidal\/}
gravitational forces, whose proper inclusion is not
encompassed by the minimal coupling framework.}
Nonetheless, it is not difficult to construct a
stress-energy tensor in the above curved space--time
which is symmetric and covariantly divergenceless,
and which, for $B(x^2)\rightarrow 1$,
reduces to the flat space--time tensor (\ref{Eq:TmunuFlat}).
We observe that this latter object
can be rewritten
in the form
\begin{equation}
T^{\mu\nu}(x)
=
4 e^2 H''(x^2)\, x^{\mu}x^{\nu} -
\eta^{\mu\nu}\left[ 4 e^2 H''(x^2)\, x^2
+6 e^2 H'(x^2)
\right].
\label{Eq:TmunuFlatCrd}
\end{equation}
One notes that the metric ansatz (\ref{Eq:Metric})
has produced an Einstein tensor (\ref{Eq:EinsteinTnsr})
having a marked structural similarity to this flat
space--time stress-energy tensor (\ref{Eq:TmunuFlatCrd}).
With a view toward making the
needed modifications to this flat-space
$T^{\mu\nu}(x)$
(these modifications must, of course, disappear
when $B(x^2)$ is put to unity,
i.e., when we go to  the flat-space limit),
the conditions under which a general tensor
having this shared structual form, namely
\begin{equation}
S^{\mu\nu}(x)\equiv
F (x^2)\, x^{\mu}x^{\nu} -
g^{\mu\nu}\left[ F(x^2)\, x^2\, [B(x^2)]^{-2}
+D(x^2)\right],
\label{Eq:SDefined}
\end{equation}
is covariantly divergenceless are now
examined.  This is essentially the covariant
generalization (in the context of
our particular ``conformally flat'' metric) of
the procedure in Eq.~(\thevanishingFlatDivergenceEq).
Taking the covariant divergence of $S^{\mu
\nu}$ yields
\begin{equation}
{S^{\mu\nu}}_{;\nu}\, =
\left [ 3 F - 6 F B B' x^2 B^{-2} -2 D' B^2 \right ] x^{\mu}.
\end{equation}
For the vanishing of the covariant divergence,
it is required that
\begin{mathletters}
\begin{equation}
D'=\frac{3}{2} F B^{-2}\left [1-2 B B' x^2 B^{-2}\right ]
\label{Eq:Dp}
\end{equation}
or
\begin{equation}
D(x^2)=\frac{3}{2} \int_{x_1^2}^{x^2} d\lambda\,[B(\lambda)]^{-2}
F(\lambda)\left[ 1- 2 \lambda \frac{B'(\lambda)}{B(\lambda)}\right].
\label{Eq:D}
\end{equation}
\end{mathletters}%
\setcounter{DEq}{\value{equation}}%
In the flat-space limit where $B(x^2)$ is
taken to unity, we can readily see
that Eqs.~(\ref{Eq:SDefined})  and (\theDEq)
reduce to analogues of the three parts of
Eq.~(\thevanishingFlatDivergenceEq).
We may also readily verify that our Einstein
tensor $G^{\mu\nu}$ of Eq.~(\ref{Eq:EinsteinTnsr})
is a tensor of the form (\ref{Eq:SDefined}), which
satisfies Eq.~(\ref{Eq:D}) with lower limit
$x_1^2 = -\infty $, because far from the light
cone (i.e.\ as $|x^2|\to \infty$), $B(x^2)\to 1$.

Having established the condition (\theDEq)
for a tensor of the form (\ref{Eq:SDefined})
to be covariantly divergenceless, one still must
note that the conversion of a purely special-relativistic
contravariant stress-energy tensor, such as
$T^{\mu\nu}(x)$ of Eq.~(\ref{Eq:TmunuFlatCrd}) above,
into a contravariant stress-energy tensor suitable for
placement on the right-hand side of the Einstein
equation, can involve, among other things, its multiplication
by the factor $g^{-\,{1\over 2}}$, where
$g\equiv -\det(g_{\mu\nu})$---see the example given in
Ref.~\cite{Weinberg2}.  Multiplication of
$T^{\mu\nu}(x)$ by $g^{-\,{1\over 2}}$ ensures invariance
of the contravariant Einstein equation under uniform
rescaling by a constant factor of the space and time
inertial coordinates of the local freely-falling reference
frames,\footnote{The inertial space and time coordinates of
the local freely-falling reference frames typically enter
implicitly into general-relativistic expressions via the
covariant metric tensor $g_{\mu\nu}$, which depends on them
in a bilinear, Lorentz-invariant manner \cite{Weinberg4}.
While the affine connection (the ``gravitational force'') and
the {\em covariant\/} Ricci and Einstein tensors are, in fact,
invariant under uniform rescaling of these inertial space and
time coordinates by a constant factor, the metric objects
$g_{\mu\nu}$, $g^{\mu\nu}$, and $g$ respectively have weights of
$+2$, $-2$, and $+8$ for such inertial rescaling.  Thus the
{\em contravariant\/} Einstein tensor $G^{\mu\nu}$ has inertial
rescaling weight $-4$, while the special-relativistic form for
$T^{\mu\nu}(x)$ given by Eq.~(\ref{Eq:TmunuFlatCrd}) has
inertial rescaling weight zero, as it obviously has no
dependence on $g_{\mu\nu}$.  Multiplication of this
special-relativistic $T^{\mu\nu}(x)$ by the factor
$g^{-\,{1\over 2}}$ gives the resultant object inertial rescaling
weight $-4$, which matches that of $G^{\mu\nu}$.}
while retaining the proper special-relativistic limiting form
for the resultant tensor as space--time becomes flat.
For the metric ansatz of Eq.~(\ref{Eq:Metric}),
$g^{-\,{1\over 2}} = B^4$, so the result of multiplying the
$T^{\mu\nu}(x)$ of Eq.~(\ref{Eq:TmunuFlatCrd}) by
$g^{-\,{1\over 2}}$ is a tensor of the form given in
Eq.~(\ref{Eq:SDefined}), with
$F(x^2)=4e^2 H''(x^2)\, [B(x^2)]^4$ and $D(x^2) =
6e^2H'(x^2)\, [B(x^2)]^2$.
To make this tensor covariantly
divergenceless, however, $D(x^2)$ must
still be modified to accord with the
prescription of Eq.~(\ref{Eq:D}).
Thus, having taken account of inertial rescaling
characteristics, the choice for the
covariantly divergenceless stress-energy tensor
$T^{\mu\nu}_B(x)$ is
\begin{mathletters}
\begin{equation}
T^{\mu\nu}_B(x) =
F(x^2)\, x^{\mu}x^{\nu} -
g^{\mu\nu}\left [F(x^2)\, x^2\, [B(x^2)]^{-2}
+D(x^2)\right ],
\label{Eq:TmunuCurved}
\end{equation}
where
\begin{equation}
F(x^2)\equiv
[B(x^2)]^4 \,
4e^2H''(x^2),
\label{Eq:FDefined}
\end{equation}
and
\begin{eqnarray}
D(x^2)\! &\equiv & \!{3\over 2}\,\int_{-\infty}^{x^2}d\lambda\,
[B(\lambda )]^{-2} F(\lambda)
\left [1- 2 \lambda \frac{B'(\lambda)}{B(\lambda)}\right]
\label{Eq:DxRaw}\\
&=& \! 6e^2\int_{-\infty}^{x^2}d\lambda\,
[B(\lambda )]^2 H''(\lambda )
\left[ 1- 2 \lambda \frac{B'(\lambda)}{B(\lambda)}\right].
\label{Eq:Dx}
\end{eqnarray}
\end{mathletters}%
\setcounter{TmunuCurvedEq}{\value{equation}}%
The $T^{\mu\nu}_B(x)$ of
Eq.~(\theTmunuCurvedEq) has inertial
rescaling weight $-4$ (see the preceding footnote), is symmetric and
covariantly divergenceless, and clearly reduces
to the flat-space $T^{\mu\nu}(x)$ of Eq.~(\ref{Eq:TmunuFlatCrd})
(and Eq.~(\ref{Eq:TmunuFlat})) when we put
$B(x^2)$ to unity.  It is also clear that $T^{\mu\nu}_B(x)$ is
entirely determined by the function  $F(x^2)$ given
in (\ref{Eq:FDefined}) above, just as the Einstein
tensor $G^{\mu\nu}$ is determined in the same manner by
its ``$F(x^2)$ part'', namely $-8B^3B''$.  So to solve the
contravariant Einstein equation
\begin{equation}
G^{\mu\nu} = -8\pi G\,T^{\mu\nu}_B,
\label{Eq:EinsteinFEQ}
\end{equation}
it is thus sufficient to equate these ``$F(x^2)$ parts'':
\begin{equation}
-8 [B(x^2)]^3 B''(x^2) =
-8\pi G
[B(x^2)]^4
4 e^2
H''(x^2).
\end{equation}
Thus one needs to solve an ordinary second-order linear
homogeneous differential equation of standard form:
\begin{equation}
B''(x^2) = 4\pi Ge^2H''(x^2)B(x^2).
\label{Eq:BLinDiff}
\end{equation}
Let us now formally express its solution as the infinite
perturbation expansion series
\begin{equation}
B(x^2) = \sum_{n = 0}^{\infty}\left (4\pi Ge^2\right )^n
b^{(n)}(x^2),
\label{Eq:BPertExp}
\end{equation}
which we substitute into Eq.~(\ref{Eq:BLinDiff}).  Using
the boundary condition that $B(x^2)\to 1$ as
$|x^2|\to\infty$, we obtain that $b^{(0)}(x^2) = 1$ and
\begin{equation}
b^{(n)}(x^2) = \int_{-\infty}^{x^2}d\lambda\,
\int_{-\infty}^{\lambda}
d\lambda '\, H''(\lambda ')\, b^{(n - 1)}(\lambda ')
\quad\hbox{for $n = 1, 2, 3,$\dots .}
\label{Eq:bRecurrence}
\end{equation}
In particular, for $n = 1$ we obtain
$b^{(1)}(x^2) = H(x^2)$.  For orders higher than the
first, we are unable to obtain exact results for
the $b^{(n)}(x^2)$.  However, so long as
$|4\pi Ge^2 H(x^2)|\ll 1$, we should be able to rely
on the results which we have obtained through first
order.  The asymptotic behaviour of $H(x^2)$ given in
Eq.~(\ref{Eq:HLightCone}) thus tells us that the
first-order perturbative results are adequate in the
region $|x^2|\gg Ge^2$.  On the other hand, in the
region $|x^2|\ll 1/m^2$, where Eq.~(\ref{Eq:HLightCone})
is the asymptotically valid representation of $H(x^2)$,
that relationship turns out to actually permit the
determination of the asymptotic forms in this region of
{\em all\/} of the $b^{(n)}(x^2)$.  Bearing in mind our results
through first order, we make the following asymptotic ansatz:
\begin{equation}
b^{(n - 1)}(x^2)\sim c_{n - 1}\left (H(x^2)\right )^{n - 1}
\sim c_{n - 1}\left (-iK\, {\ln (m^2x^2)\over x^2}\right )^
{n - 1}\quad\hbox{for $|x^2|\ll 1/m^2$,}
\label{Eq:bAnsatz}
\end{equation}
where, of course, $c_0 = 1$ and, from Eq.~(\ref{Eq:HLightCone}),
$K = 1/(12\pi^3)$.  Since we readily see that
\begin{equation}
H''(x^2)\sim -2iK\, {\ln (m^2x^2)\over (x^2)^3}
\quad\hbox{for $|x^2|\ll 1/m^2$,}
\label{Eq:HppAsymp}
\end{equation}
we calculate from Eq.~(\ref{Eq:bRecurrence}) that
\begin{equation}
b^{(n)}(x^2)\sim {2c_{n - 1}\over n(n + 1)}
\left (-iK\, {\ln (m^2x^2)\over x^2}\right )^n
\sim {2c_{n - 1}\over n(n + 1)}\left (H(x^2)\right )^n
\quad\hbox{for $|x^2|\ll 1/m^2$.}
\label{Eq:bInduced}
\end{equation}
Thus we have obtained the recurrence relation for the
coefficients $c_n$ of our ansatz:
\begin{equation}
c_n = {2c_{n - 1}\over n(n + 1)},
\label{Eq:cRecurrence}
\end{equation}
which, for our known zeroth-order $c_0 = 1$, properly yields
our known first-order $c_1 = 1$, and may readily be solved
for all orders:
\begin{equation}
c_n = {2^n\over n!\, (n + 1)!}\quad
\hbox{for $n = 0, 1, 2,$\dots .}
\label{Eq:cFormula}
\end{equation}
Putting this result and our ansatz of Eq.~(\ref{Eq:bAnsatz})
into the perturbative expansion series of
Eq.~(\ref{Eq:BPertExp}), we obtain
\begin{equation}
B(x^2)\sim \sum_{n = 0}^{\infty}{\left (8\pi Ge^2 H(x^2)
\right )^n\over n!\, (n + 1)!} =
{I_1(\, 2\sqrt{8\pi Ge^2H(x^2)\, }\, )\over
\sqrt{8\pi Ge^2H(x^2)\, }}\quad\hbox{for $|x^2|\ll 1/m^2$,}
\label{Eq:BofHSummed}
\end{equation}
where we have been able to actually sum, in terms of
the modified Bessel function $I_1$, the perturbation
expansion in the region $|x^2|\ll 1/m^2$.
Eq.~(\ref{Eq:BofHSummed}) clearly incorporates the
unadulterated perturbative result through first
order, which, as we have discussed above, ensures this
equation's accuracy so long as $|x^2|\gg Ge^2$.  In view
of this equation's validity for $|x^2|\ll 1/m^2$
{\em as well}, plus the fact that
$Ge^2m^2\approx 1.3\times 10^{-47}$, we conclude that
\begin{equation}
B(x^2)\approx {I_1(\, 2\sqrt{8\pi Ge^2H(x^2)\, }\, )\over
\sqrt{8\pi Ge^2H(x^2)\, }}\quad\hbox{uniformly in $x^2$.}
\label{Eq:BofHFound}
\end{equation}
The complex nature of
$H(x^2)$ of course causes $B(x^2)$ to be
complex as well---our metric is a
virtual one, as was mentioned at
the beginning of this Section.

The desired virtual metric
$g_{\mu\nu} = [B(x^2)]^{-2}\eta_{\mu\nu}$ is now
in hand from the result in
Eq.~(\ref{Eq:BofHFound}).
{}From this virtual metric we obtain the
flat-space (Minkowskian) virtual gravitational
field tensor $h_{\mu\nu}$:
\begin{equation}
h_{\mu\nu} = g_{\mu\nu} - \eta_{\mu\nu} =
\left ([B(x^2)]^{-2} - 1\right ) \eta_{\mu\nu}.
\label{Eq:graviton}
\end{equation}
For our virtual Einstein tensor $G^{\mu\nu}$ of
Eq.~(\ref{Eq:EinsteinTnsr}), the portion that is
linear in $h_{\mu\nu}$ is given by
\begin{mathletters}
\begin{eqnarray}
G^{(1)\,\mu\nu} \! &=& \!x^{\mu}x^{\nu}\left (4(B^{-2}-1)''\right )
-\eta^{\mu\nu}
\left (4x^2(B^{-2} - 1)'' + 6(B^{-2} - 1)'\right )
\label{Eq:Einstein1Raw}\\
&=& \! \left (\partial^{\mu}\partial^{\nu} - \eta^{\mu\nu}\partial_{\alpha}%
\partial^{\alpha}\right )\left ([B(x^2)]^{-2} - 1\right ).
\label{Eq:Einstein1}
\end{eqnarray}
\end{mathletters}%
Of course,
$G^{(1)\,\mu\nu}$ is
divergenceless in the ordinary flat-space
(Minkowskian) sense, so it is customary
to define the flat-space
virtual stress-energy tensor which {\em includes\/}
the virtual gravitational effects as \cite{Weinberg}:
\begin{equation}
\tau^{\mu\nu}\equiv -\,{1\over 8\pi G}\, G^{(1)\,\mu\nu}.
\label{Eq:tauDef}
\end{equation}
Thus $\tau^{\mu\nu}$ is the self-gravitationally
corrected flat-space
virtual stress-energy tensor $T^{\mu\nu}_G (x)$
which properly replaces our
``$G = 0$'' $\> T^{\mu\nu}(x)$ of
Eqs.~(\ref{Eq:TmunuFlat}) and
(\ref{Eq:TmunuFlatCrd}).
Using Eqs.~(\ref{Eq:Einstein1})
and (\ref{Eq:tauDef}), we write it in the form
\begin{mathletters}
\begin{eqnarray}
T^{\mu\nu}_G (x) \! &=&\! e^2
\left (\partial^{\mu}\partial^{\nu} - \eta^{\mu\nu}\partial_{\alpha}
\partial^{\alpha}\right )\left (-\,{1\over 8\pi Ge^2}\right )
\left ([B(x^2)]^{-2} - 1\right ),
\label{Eq:TGRaw}\\
&=& \! e^2\left (\partial^{\mu}\partial^{\nu} - \eta^{\mu\nu}\partial_{\alpha}
\partial^{\alpha}\right )H_G(x^2),
\label{Eq:TG}
\end{eqnarray}
where
\begin{equation}
H_G(x^2)\equiv -\,{1\over 8\pi Ge^2}\,\left ([B(x^2)]^{-2} - 1\right ).
\label{Eq:HGDef}
\end{equation}
\end{mathletters}%
If we further define
\begin{equation}
h_G(x^2)\equiv \partial_{\alpha}\partial^{\alpha}H_G(x^2),
\label{Eq:hGDef}
\end{equation}
we have in hand all the ingredients needed
for the discussion of the
self-gravitationally corrected version of
the Feynman diagram of Fig.~\ref{Fig:VacPol}.

\section{Discussion}
\label{Sect:Discussion}

For the self-gravitationally
corrected case, we still
have relations analogous to those of Eqs.~(\theFlatTmunuEq):
\begin{mathletters}
\begin{equation}
T^{\mu\nu}_G (x) = e^2\,\left(\partial^{\mu}\partial^{\nu} -
\eta^{\mu\nu}\partial_{\alpha}\partial^{\alpha}\right)H_G(x^2)
\label{Eq:RelationTcorr}
\end{equation}
and
\begin{equation}
\partial_{\alpha}\partial^{\alpha} H_G(x^2) = h_G(x^2),
\label{Eq:RelationHcorr}
\end{equation}
\end{mathletters}%
\setcounter{selfcorrectedEq}{\value{equation}}%
where, as shown in Section~\ref{Sect:SelfGravityCorrection},
$H_G(x^2)$ is simply related to a
dimensionless virtual metric function $B(x^2)$:
\begin{mathletters}
\begin{equation}
H_G(x^2) = -\,{1\over 8\pi G e^2}\,\left([B(x^2)]^{-2} - 1\right),
\label{Eq:HGofB}
\end{equation}
and this virtual metric function $B(x^2)$ is,
in turn, related to $H(x^2)$:
\begin{equation}
B(x^2)\approx {I_1(\, 2\sqrt{8\pi Ge^2H(x^2)\, }\, )\over
\sqrt{8\pi Ge^2H(x^2)\, }}\, .
\label{Eq:BofH}
\end{equation}
\end{mathletters}%
\setcounter{HGandBEq}{\value{equation}}%
Note that our virtual metric function $B(x^2)$
tends toward unity far from the
light cone ($|x^2|\to \infty$),
as we can see from Eqs.~(\ref{Eq:HPlusInfinity})
and (\ref{Eq:HMinusInfinity}) that $H(x^2)$
vanishes in that limit.
Equations (\ref{Eq:RelationHcorr}) and (\ref{Eq:HGofB})
also yield
\begin{equation}
h_G(x^2) =
-\,{\partial_{\alpha}\partial^{\alpha}[B(x^2)]^{-2}\over
8\pi G e^2}
\, .
\label{Eq:hGsolution}
\end{equation}
{}From Eqs.~(\theHGandBEq) and (\ref{Eq:hGsolution}),
we can deduce that to zeroth order in $G$, $H_G(x^2)$
and $h_G(x^2)$ reduce to $H(x^2)$
and $h(x^2)$ respectively (and thus, from
Eq.~(\ref{Eq:RelationTcorr}),
$T^{\mu\nu}_G (x)$ reduces to $T^{\mu\nu}(x)$), as must be
true for physical consistency.
However, very near to the light cone, the
objects $H_G(x^2)$, $h_G(x^2)$,
and $T^{\mu\nu}_G (x)$ behave very
differently from their counterparts $H(x^2)$,
$h(x^2)$, and $T^{\mu\nu}(x)$
respectively---which do not take
account of the self-gravitational
correction.  We know from Eq.~(\ref{Eq:HLightCone}),
that $H(x^2)\sim -iK\ln (m^2x^2)
/x^2$ for $|x^2|\ll 1/m^2$.  Thus, when
$|x^2|\lesssim Ge^2$, $|8\pi Ge^2H(x^2)|\gg 1$, and
we may replace the modified Bessel function
$I_1$ of Eq.~(\ref{Eq:BofH}) by its large-argument
asymptotic form:
\begin{mathletters}
\begin{equation}
B(x^2)\sim \left (1\over 4\pi\right )^{1\over 2}
{\exp(\, 2\sqrt{8\pi Ge^2H(x^2)\, }\, )
\over [8\pi Ge^2H(x^2)]^{3\over 4}}
\quad\hbox{for $|x^2|\lesssim Ge^2$.}
\label{Eq:Blighcone}
\end{equation}
The above asymptotic form is in essential agreement
with the result of applying the WKB approximation
directly to our original differential equation
(\ref{Eq:BLinDiff}) in this region.
In conjunction with Eq.~(\ref{Eq:hGsolution}),
it gives the asymptotic light cone behaviour for $h_G(x^2)$:
\begin{equation}
h_G(x^2)\sim h(x^2)\,4\ln(m^2x^2)[B(x^2)]^{-2}
\quad\hbox{for $|x^2|\lesssim Ge^2$.}
\label{Eq:hGlightcone}
\end{equation}
\end{mathletters}%
\setcounter{hGandBLightConeEq}{\value{equation}}%
The asymptotic light-cone behaviour of $h(x^2)$
as given by Eq.~(\ref{Eq:hLightCone}),
namely $h(x^2)\sim 4iK(x^2)^{-2}$
as $x^2\to 0$, is non-integrably singular,
which is a far cry from that of $h_G(x^2)$ as
given above, with its factor of $[B(x^2)]^{-2}$.
The essential singularity on the light cone,
which Eq.~(\ref{Eq:Blighcone}) reveals in the
behaviour of $B(x^2)$, plays the critical r\^ole in
Eq.~(\ref{Eq:hGlightcone}), with its factor
of $[B(x^2)]^{-2}$, of driving $h_G(x^2)$ very
decisively to zero as $x^2\to 0$.
Thus, unlike the
logarithmically divergent $\hat h(q^2)$,
$\hat h_G(q^2)$ is fully convergent and finite.

{}From Eqs.~(\theselfcorrectedEq) we can show
that the self-gravitationally corrected objects
${\widehat T_G}^{\mu\nu}(q)$ and
$\hat h_G(q^2)$ satisfy the analogue of
Eq.~(\ref{Eq:That}),
\begin{equation}
{\widehat T_G}^{\mu\nu}(q) = e^2\,\left(\eta^{\mu\nu}q^2
- q^{\mu}q^{\nu}
\right) \left(-\,{\hat h_G(q^2)\over q^2
+ i\epsilon}\right).
\label{Eq:Thatanalog}
\end{equation}
We observe that ${\widehat T_G}^{\mu\nu}(q)$ is not,
strictly speaking, second-order in $e$, since $\hat h_G(q^2)$
depends on {\em all\/} orders of $Ge^2$.  This is in accord
with Haag's Theorem, which maintains that there can exist no
strictly perturbative development in powers of $e$ of quantum
electrodynamics.  Eq.~(\ref{Eq:Thatanalog}) may be reexpressed
in a form analogous to Eq.~(\ref{Eq:Tsplih}),
\begin{equation}
{\widehat T_G}^{\mu\nu}(q) = -\,{e^2\over q^2}\,\left(
\eta^{\mu\nu}q^2 - q^{\mu}q^{\nu}\right )\,
\left (\left. \hat h_G\right|_{q^2 = 0} + \left (
\hat h_G(q^2) - \left. \hat h_G\right|_{q^2 = 0}\right )\right ).
\label{Eq:Tsplihanalog}
\end{equation}
In Eq.~(\ref{Eq:Tsplihanalog}),
however, $\hat h_G|_{q^2 = 0}$ is
not logarithmically
divergent, but fully finite, and it
satisfies the well-defined integral
relationship
\begin{equation}
\left. \hat h_G\right|_{q^2 = 0} = \int d^4x\,h_G(x^2).
\label{Eq:IntegralRelationRaw}
\end{equation}
Putting Eq.~(\ref{Eq:RelationHcorr})
into Eq.~(\ref{Eq:IntegralRelationRaw}), we arrive at
\begin{equation}
\left. \hat h_G\right|_{q^2 = 0} = \int d^4x\,
\left(\partial_\alpha\partial^\alpha
H_G(x^2)\right ),
\label{Eq:IntegralRelation}
\end{equation}
where the integral on the right-hand
side is finite and well-defined.
As it is the well-defined integral
of a perfect differential, it must
vanish (provided, of course,
that the large $|x^2|$ asymptotic behaviour
is appropriate---it is clear from
Eq.~(\theHGandBEq) that the
large $|x^2|$ asymptotic behaviour of $H_G(x^2)$ is the
same as that of $H(x^2)$, which, as
we see from Eqs.~(\ref{Eq:HPlusInfinity}) and
(\ref{Eq:HMinusInfinity}), tends toward zero for large $|x^2|$).

The vanishing of $\hat h_G|_{q^2 = 0}$
means that this insertion
$\widehat T_{G}^{\mu\nu}(q)$,
once gravitationally corrected, does
{\em not\/}
contribute to charge renormalization {\em at all\/},
removing the paradox of
an ultraviolet-dominated
(not to mention infinite) object influencing
the scale of the extreme low-energy
scattering strength.  Thus, in
the self-gravitationally corrected case,
the name ``vacuum polarization''
for this diagram is seen to be a serious
misnomer.  In fact, the nature of
the diagram as represented in
Fig.~\ref{Fig:VacPol} makes it very clear that
the production of the virtual pair is
dependent {\em only\/} on the presence of
the virtual photon---indeed,
that virtual photon {\em dissociates\/} into
the virtual pair, which subsequently recombines.
Arising entirely from
the virtual photon, the transient
existence of this {\em single\/} virtual
pair could easily be {\em far removed in
space--time from any other charged
particles\/},
so it can hardly be expected to systematically
shield (renormalize) their charges.
Of course, if the probability to make
such a virtual pair somehow were {\em divergent\/}
throughout space--time, we
could no doubt have
overwhelming shielding of other charges---that is
presumably the diseased state of
affairs before the self-gravitational
correction of $T^{\mu\nu}(x)$ is taken into account.
In view of the ``vacuum
polarization'' diagram's
reasonable lack of any contribution
whatever to charge renormalization
(once self-gravitationally corrected), the
``vacuum polarization'' misnomer
accorded it needs to be discarded and
replaced by a more physically accurate
shorthand nomenclature---in view of
our discussion at the end of
Section~\ref{Sect:VPConfiguration},
``vacuum stress-energy'' would seem
to be an appropriate replacement prefix.

If, for a moment, $\hat h_G(q^2)$ is regarded to be merely a
particular ultraviolet cutoff of the logarithmically divergent
$\hat h(q^2)$, with $G$ playing the r\^ ole of cutoff parameter,
it can be said that a cutoff technique has been found for
which the diagram's contribution to charge renormalization vanishes
identically over the entire range of values for the cutoff
parameter except the non-cutoff
limit value ($G = 0$ in this instance).  Having thus
encountered a cutoff method which has this property, it is clear
that one can now readily construct many others.  Given the existence
of myriad cutoff techniques which yield zero charge renormalization
contribution over their full range of cutoff parameter values
(aside from the non-cutoff limit value),
it becomes obvious that the ``need''
which this ``vacuum polarization'' diagram poses to effect charge
renormalizaton is, in fact, a chimera.

Until one gets extremely close to the
light cone ($|x^2| \approx G e^2$),
$h(x^2)$ remains an excellent
approximation to $h_G(x^2)$.  In particular,
for $1/m^2\gg |x^2|\gg G e^2$,
$h_G(x^2) \approx 4iK (x^2)^{-2}$.
Thus, for the ``normal'' high-momentum-transfer range, $m^2%
\ll |q^2|\ll (Ge^2)^{-1}$, we can expect
$\hat h_G(q^2)$ in Eq.~(\ref{Eq:Thatanalog}) to
behave logarithmically in
$q^2$, $\hat h_G(q^2)\sim C_1 \ln (|q^2|/m^2)$,
the familiar logarithmic growth with
momentum transfer of the effective
coupling at high momentum transfer \cite{BjorkenDrell3}.
However, once we reach the
ultrahigh momentum transfer regime,
$|q^2|\gtrsim (Ge^2)^{-1}$,
Eqs.~(\thehGandBLightConeEq)
for the asymptotic behaviour of
$h_G(x^2)$ very close to the light cone
imply that
\begin{equation}
|\hat h_G(q^2)|\sim C_2\>
[\, \ln(\, |q^2|/m^2)\, ]^{5\over 2}\;\,
[\, G e^2 |q^2|\, ]^{3\over 2}\>
\exp(-8\sqrt{K\pi G e^2 |q^2|
\ln(\, |q^2|/m^2)\, }\> )\, .
\label{Eq:hGlightconeq}
\end{equation}
Thus the logarithmic growth of
effective coupling with momentum
transfer,
after reaching a
peak value of approximately $e^2
(1 + (e^2/(3\pi )) \ln ((Ge^2m^2)^{-1}))$
at $|q^2|\approx (Ge^2)^{-1}$,
ultimately
proceeds to abruptly
collapse back to $e^2$ for larger $|q^2|$.
Self-gravitational effects
enforce ultimate ``asymptotic freedom'' even in
quantum electrodynamics!
Of course these considerations concerning
such a ``self-gravitational
form factor'' in quantum electrodynamics
are of purely academic
interest---momentum transfers of order $(Ge^2)^{-1}$
are utterly unattainable.
In that vein, however, it is
nonetheless fascinating that the
{\em peak\/} effective coupling strength
of around $e^2 (1 + (e^2/(3\pi ))
\ln ((Ge^2m^2)^{-1}))$ is still very
much in the vicinity of $e^2$.
The portentous ``logarithmic rise in
effective coupling strength'' never
can amount to a great deal, thanks
to the limiting effect of self-gravitation.
This example suggests in
particular that the idea that the
effective electromagnetic coupling
strength may become equal to
that of the strong interactions at
sufficiently high energy may not be a reasonable one.

The ingredients needed for the evaluation of
the self-gravitationally corrected version of
the quartically divergent second-order
electromagnetic vacuum-to-vacuum amplitude
correction Feynman diagram shown in
Fig.~\ref{Fig:VacVac} are also in hand.
This diagram only differs
from that of Fig.~\ref{Fig:VacPol} in that the
single photon propagator now has each of its
two ends attached respectively to one of the two
vertices of the virtual electron--positron loop.
Indeed, the value of the self-gravitationally
corrected version of this diagram is simply
\begin{equation}
\int{d^4q\over (2\pi )^4}\>\eta_{\mu\nu}\widehat T_G^{\mu\nu}(q) =
-3 e^2 \int{d^4q\over (2\pi )^4}\>\hat h_G(q^2) =
-3 e^2 \left . h_G\right |_{x^2 = 0} = 0,
\label{Eq:VacVacCorr}
\end{equation}
where we have made use of Eq.~(\ref{Eq:Thatanalog}),
of the fact that $h_G(x^2)$ is the four-dimensional
Fourier transform of $\hat h_G(q^2)$, and of
the vanishing of $h_G|_{x^2 = 0}$ which follows
from Eqs.~(\thehGandBLightConeEq).  Without the
self-gravitational correction, of course
$h(x^2) \sim 4iK(x^2)^{-2}$ as $x^2 \to 0$,
which accords with this diagram's quartic
divergence.  As we have mentioned in
Section~\ref{Sect:Introduction}, the vanishing
of this diagram upon its self-gravitational
correction obviates the need for the usual
fiat injunction that such a ``disconnected''
diagram is to be ``discarded'', notwithstanding
its strongly infinite value.

The ``gravity-modified'' result of Ref.~\cite{ISS}
for the ``vacuum polarization'' diagram differs
from that obtained here---the Ref.~\cite{ISS}
result does not deviate from the usual quantum
electrodynamics result aside from having a finite
charge renormalization (at least this is so if
an accompanying gauge-non-invariant term that
emerges is disregarded).  The dominant
$\ln (|q^2|/m^2)$ behaviour of the Ref.~\cite{ISS}
result as $|q^2| \to \infty$ implies that there
can be {\em no amelioration\/} of the quartic
ultraviolet divergence of the closely related
second-order vacuum-to-vacuum amplitude correction
diagram---in stark contrast to the vanishing
result obtained in Eq.~(\ref{Eq:VacVacCorr}) above
for that diagram.  The ``graviton superpropagator''
approximation of Ref.~\cite{ISS} also produces
gauge invariance difficulties and is not of Haag's
Theorem character.

Finally, it must be emphasized that although the
careful propagation of gravitons with the full
Einstein equation that has been pursued in this
paper is technically challenging, the usual
approximations which split off the non-linear
interactions of gravitons {\em with each other\/}
to be subsequently treated {\em perturbatively\/}
are disastrously inappropriate in the extreme
ultraviolet regime, where those interactions in
fact dominate the gravitational physics and
produce ``black-hole-like'' damping phenomena.
The {\em perturbative\/} mistreatment in the extreme
ultraviolet of the there {\em dominant\/} non-linear
terms produces physically and mathematically
nonsensical results, namely the deluge of
ultraviolet divergences which render what is
usually termed ``quantized gravity theory''
unrenormalizable.  The actual physical
character of gravitation in the extreme
ultraviolet clearly lies at the opposite
pole from these perturbative
artifacts---gravitation manifests an
overwhelming tendency to damp, not
diverge, in that limit,
as its ``black-hole-like'' aspects come
into their own.  The usual intractable
ultraviolet divergences of quantized
gravity theory are thus clearly seen
to have nothing to do with the actual
physics, but everything to do with
unthinking utilization of an
extraordinarily ill-suited perturbation
approach.

\subsection*{Acknowledgments}

The author is profoundly indebted to Harald H. Soleng for
intensive discussions during the evolution of this work,
as well as recommendation of references and checking of
the Einstein equation.  Professors Finn Ravndal,
Steven Giddings and Richard Woodard also made suggestions
and critical comments which motivated the clarification
and extension of the physical ideas of the paper.

\newpage

%              REFERENCES

\newpage
\subsection*{Figure captions}
\vspace{2.0cm}
\begin{figure}[h]
  \caption{The second-order ``vacuum polarization'' radiative
               correction insertion Feynman diagram.}
  \label{Fig:VacPol}
\end{figure}

\vspace{2.0cm}

\begin{figure}[h]
  \caption{The second-order vacuum-to-vacuum amplitude
               correction Feynman diagram.}
  \label{Fig:VacVac}
\end{figure}

\end{document}